\documentclass[10pt,conference]{IEEEtran}
\IEEEoverridecommandlockouts
\usepackage{cite}
\usepackage{amsmath,amssymb,amsfonts}
\usepackage{algorithmic}
\usepackage{graphicx}
\usepackage{textcomp}
\usepackage{xcolor}
\usepackage[bookmarks=true,breaklinks=true,colorlinks,citecolor=blue,linkcolor=blue,urlcolor=blue]{hyperref}
\usepackage{microtype}[final]
\usepackage{quantikz}
\usepackage{tikz}

\makeatletter
\let\MYcaption\@makecaption
\makeatother

\usepackage[labelformat=simple]{subcaption}

\makeatletter
\let\@makecaption\MYcaption
\makeatother

\def\BibTeX{{\rm B\kern-.05em{\sc i\kern-.025em b}\kern-.08em
    T\kern-.1667em\lower.7ex\hbox{E}\kern-.125emX}}
\begin{document}

\title{Understanding Side-Channel Vulnerabilities in Superconducting Qubit Readout Architectures}

\author{
\IEEEauthorblockN{Satvik Maurya$^1$, Chaithanya Naik Mude$^1$, Benjamin Lienhard$^{2,3}$, Swamit Tannu$^1$}
\IEEEauthorblockA{$^1$Department of Computer Sciences, University of Wisconsin-Madison, Madison, WI 53706 USA}
\IEEEauthorblockA{$^2$Department of Chemistry, Princeton University, Princeton, NJ 08544 USA}
\IEEEauthorblockA{$^3$Department of Electrical and Computer Engineering, Princeton University, Princeton, NJ 08544 USA}

\thanks{B.L. is supported by the Swiss National Science Foundation (Postdoc.Mobility Fellowship grant \#P500PT\_211060).}
}


\maketitle

\begin{abstract}
Frequency-multiplexing is an effective method to achieve resource-efficient superconducting qubit readout. Allowing multiple resonators to share a common feedline, the number of cables and passive components involved in the readout of a qubit can be drastically reduced. However, this improvement in scalability comes at the price of a crucial non-ideality -- an increased readout crosstalk. Prior works have targeted building better devices and discriminators to reduce its effects, as readout-crosstalk-induced qubit measurement errors are detrimental to the reliability of a quantum computer. However, in this work, we show that beyond the reliability of a system, readout crosstalk can introduce vulnerabilities in a system being shared among multiple users. These vulnerabilities are directly related to correlated errors due to readout crosstalk. These correlated errors can be exploited by nefarious attackers to predict the state of the victim qubits, resulting in information leakage.

\end{abstract}

\begin{IEEEkeywords}
Frequency-multiplexed readout, superconducting qubits, readout crosstalk, side-channel attacks.

\end{IEEEkeywords}

\section{Introduction} 


Quantum processors typically demand cryogenic or ultra-high vacuum environments to operate long-coherence quantum bits (qubits) with high-fidelity control and readout. Additionally, they rely on complex and expensive control and readout infrastructure, often necessitating frequent maintenance checks by specialized personnel. These demands present considerable obstacles to deploying quantum hardware on premises. The quantum cloud computing paradigm offers a solution by consolidating quantum computers and granting access via software Application Programming Interfaces (API). This approach democratizes access to quantum computing, fostering the growth of quantum software and applications. 


Currently, the prevailing model for quantum cloud services is time-sharing, granting a single user exclusive access to the quantum computer for a designated period. However, this approach faces limitations due to the under-utilization of the quantum processor amidst high demand. The main issue stems from the limited gate fidelity of current quantum systems, which confines qubit entanglement to a small subset, thereby restricting the size of quantum circuits to just a few tens of qubits. Consequently, a significant portion of qubits remain unused on most quantum processors. This inefficiency underscores a critical misalignment between the supply and demand for quantum computing resources. Despite the scarce availability of operational quantum computers and substantial demand, the prevailing time-sharing model fails to maximize utilization, even on the most advanced machines. 

To reduce under-utilization, cloud vendors can opt for a multi-tenant model, partitioning qubits and distributing among multiple users to enable concurrent access to improve hardware utilization~\cite{das2019case}. Most jobs on the quantum cloud can be serviced concurrently, as only a fraction of the qubits are used per job. Therefore, the multiprogramming model has the potential to significantly improve the queuing delay problem on the quantum cloud~\cite{Ravi2021}. However, the multiprogramming model can lead to information leakage, where one user learns the other user's computations. In a cloud setting, such information leakage in classical computers has been demonstrated and used to steal cryptographic keys and other sensitive information~\cite{brumley2005remote,side1,side2,side3,side4}. Recently, multiple studies have shown denial-of-service attacks and side channels on multi-tenant quantum computers. 

\begin{figure}[t]
    \centering
\includegraphics[width=\linewidth]{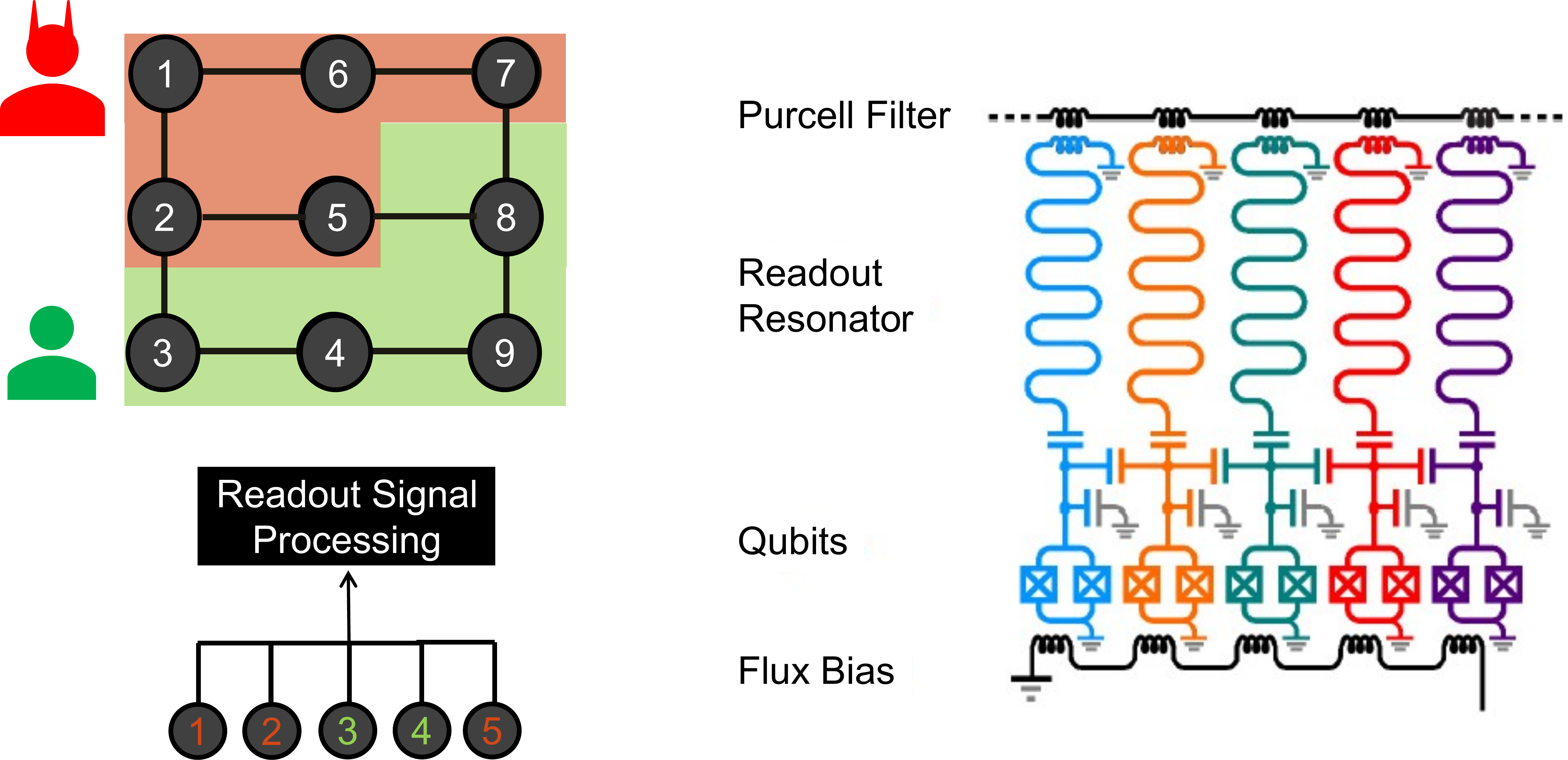}
    \caption{The left panel depicts a graphical representation of a side-channel attack on a nine-qubit quantum processor utilizing readout crosstalk. On the right panel is a circuit schematic featuring five superconducting transmon qubits~\cite{Lienhard2022}. In this schematic, the qubit transition frequencies are controlled via a global flux bias. Each qubit is capacitively linked to a quarter-wave readout resonator, which is inductively coupled to a bandpass (Purcell) filtered feedline.}
    \label{fig:intro}
    \vspace{-0.1in}
\end{figure}


Here, we investigate the potential security risks in superconducting quantum cloud computers due to information leakage between users enabled by readout crosstalk. The physical qubits are divided among users in environments where quantum computers are shared (a multi-tenant model). This division is carefully planned to allow users to run their quantum circuits efficiently, be exposed to minimal noise, and, ideally, be free from interference between the different users' activities. Typically, qubits are distributed among them according to a connectivity graph. This distribution aims to ensure each user operates on a connected subset of qubits, optimizing their circuit's performance by reducing the need for SWAP operations, which can introduce noise. As shown in Fig.~\ref{fig:intro}, one user might be assigned qubits 3, 4, 9, and 8, while another potentially malicious user could have qubits 1, 2, 5, 6, and 7. Although users are supposed to be isolated at the software level, the hardware might connect qubits and the processing of quantum information in ways that can unintentionally link users' data.

At larger scales, this issue becomes increasingly critical. For example, frequency-multiplexed readout architectures are employed to enhance scalability and simplify wiring in superconducting qubit architectures, where multiple qubits are linked to a single feedline for qubit-state readout. This configuration decreases the required complexity and equipment, such as Purcell filters regulating the qubits' environment. However, this study underscores a potential drawback: if a ``victim'' user's qubits share the same feedline as a ``malicious'' user's qubits, the latter could exploit this connection to access information they should not have. 


We analyze a dataset of one-microsecond-long readout traces obtained from a five-qubit superconducting quantum computer, as described in Reference~\cite{Lienhard2022}, to investigate the potential information leakage through readout crosstalk. Five qubits are frequency multiplexed in this hardware setup and linked to a single feedline. This dataset has been previously utilized in a study by Maurya et al.~\cite{Maurya2023} to assess the effectiveness of hardware-friendly readout discriminators. It comprises 50,000 readout traces for each of the 32 state configurations representing the five qubits. The dataset is balanced in each qubit's states prepared in the $\ket{0}$ and $\ket{1}$ state. 

We characterize readout crosstalk to understand the potential state-dependence of crosstalk between target and spectator qubits. As such, crosstalk can lead to significant sources of side-channel leakage. To illustrate this, consider the example wherein a chain of five qubits is shown below: A1 and A2 are attacker qubits controlled by the malicious user, and V1, V2, and V3 are qubits controlled by the victim user.  

\begin{center}
\vspace{0.1in}
\begin{tikzpicture}
    \node[draw, circle,minimum size=0.15cm] (A1) at (0,0) {A1};
    \node[draw, circle,minimum size=0.15cm] (V1) at (1.5,0) {V1};
    \node[draw, circle,minimum size=0.15cm] (V2) at (3,0) {V2};
    \node[draw, circle,minimum size=0.15cm] (V3) at (4.5,0) {V3};
    \node[draw, circle,minimum size=0.15cm] (A2) at (6,0) {A2};
    
    \draw (A1) -- (V1);
    \draw (V1) -- (V2);
    \draw (V2) -- (V3);
    \draw (V3) -- (A2);

\end{tikzpicture}
\vspace{0.1in}
\end{center}

The attacker wants to estimate the state of V1, V2, and V3 by preparing A1 and A2 in specific states and measuring them. Such an attack is possible if the readout crosstalk, thereby, the crosstalk-induced errors on A1 and A2 modulate with V1, V2, and V3 states. Therefore, we investigate if the average readout crosstalk experienced by a physical qubit fluctuates due to the different states of spectator qubits. 

Analyzing the experimental readout trace dataset shows that for several combinations of attacker and victim qubit mappings, the $\ket{0} \rightarrow\ket{1}$ bitflip probability is influenced by the state of victim qubits. We hypothesize the bitflip is caused by the crosstalk during the readout process and, thereby, can be used as a side channel. We train a Support Vector Machine (SVM) classifier to predict the output state of victim qubits using the readout outcomes of the attacker qubits. We observe certain attack configurations. The attacker can predict the output state of victim qubits with more than 80\%, indicating high informational leakage.
Furthermore, classification accuracy increases information leakage as we improve readout discrimination. For example, compared to a simple readout discrimination algorithm that uses a Matched Filter~\cite{PhysRevA.91.022118}, a sophisticated readout algorithm like HERQULES~\cite{Maurya2023} leaks more information. The HERQULES discrimination approach uses a relaxation-matched filter to detect readout errors due to state transitions and a Neural Network to account for readout crosstalk.

Quantum hardware is both costly and scarce. To promote innovation, it is crucial to democratize access by sharing this hardware efficiently among various users while securing confidential computations. This paper takes a step toward understanding the side channel leakage caused by readout crosstalk and discusses potential strategies to defend against such attacks.  
\section{Background and Motivation} 

\subsection{Security of Quantum Computers}


Despite their current limitations, quantum computers find applications in various fields, including finance, cryptography, and quantum simulations~\cite{brooksNature2023QuantumComputing}. Researchers often seek real-world problem instances, which may contain sensitive data, for exploration. Moreover, the hardware and software primitives developed for quantum computing hold considerable intellectual property value, and the outcomes generated by quantum circuits are essential. Additionally, the components designed for quantum computing can be repurposed for applications in quantum networking, such as Quantum Key Distribution (QKD). Therefore, it is crucial to understand the risks associated with sharing quantum resources among multiple users, particularly regarding side channels. Side channel information leakage refers to the inadvertent disclosure of sensitive data through indirect means, such as data-dependent fluctuations in power or execution time. Attackers can exploit these side channels to infer the quantum circuit being executed or the data being generated on the quantum computer.

Alongside the imperative of shared processors, securing quantum repeaters for quantum communication or a quantum internet presents a significant challenge. The presence of multiple quantum sensors or qubits surrounding a repeater can lead to problematic crosstalk and potential information loss. While the severity of the challenge posed by a shared readout feedline in a quantum processor, as studied here, is undoubtedly more significant, the insights gained from addressing this issue may have implications for safeguarding quantum repeaters against attacks orchestrated by malicious quantum sensors surrounding them. 




\subsection{Related Work}

Gate and readout crosstalk have prompted extensive research, focusing primarily on understanding the sources and mitigating these errors~\cite{10.1145/3373376.3378477, 10.1145/3370748.3406570}. However, these studies mainly pertain to reliability issues and have yet to be studied from a system security perspective.


Recent investigations have explored potential threats that exploit vulnerabilities in quantum computers. For instance, qubit sensing has been examined on the feasibility of recovering the state of a qubit after operation~\cite{saki2021qubit}. Another study proposes a framework for detecting malicious quantum circuits by comparing them against a database of known harmful sequences~\cite{10133711}. Additionally, a recent paper has introduced the concept of a power side-channel attack, leveraging the pulse signatures of quantum gates to infer the operations performed by a circuit, potentially identifying the entire sequence~\cite{erata2024quantum}. These studies underscore the importance of understanding the threats to data confidentiality on quantum computing platforms.

\section{Information Leakage due to Readout Crosstalk} 

\subsection{Readout Crosstalk}

In systems with multiple superconducting qubits, readout crosstalk manifests as a combination of factors: (1) interactions between the readout probe signals, (2) photon population due to residual coupling to a probe tone or neighboring readout resonators, (3) coupling between readout resonator and neighboring qubits, and (4) interactions between reflected/transmitted readout signals in the amplifier chain, mixers, or during analog demodulation and digitization.

Fast readout, essential for ancilla qubits in quantum error correction protocols, necessitates wide resonator linewidths. However, the frequency spacing between readout resonators is limited by factors such as the qubit transition frequency, the number of frequency-multiplexed probe tones, and the readout amplifier chain bandwidth. Readout crosstalk increases with the spectral overlap between resonators, implying that wider resonator linewidths result in more pronounced readout crosstalk. Consequently, readout crosstalk is a significant error source, particularly for fast frequency-multiplexed ancilla qubit readout.


\subsection{Threat Model}
We posit that an attacker's primary aim is to leak critical data generated by the target quantum circuit, thereby breaching its confidentiality.

In our threat model, we treat the attacker as having the same access level as a regular user, showcasing the attack's effectiveness even with minimal attacker privileges. We envision a scenario where the attacker and the victim operate on the same quantum computer but use separate qubit sets, with the attacker having no direct influence over the victim's circuit operations. This approach aims to develop an attack methodology that remains viable within realistic limitations, allowing attackers to extract sensitive information without direct interference in a shared cloud environment. Here, any user, including the attacker, can control a subset of qubits to run their quantum programs. This suggests that the attacker's capabilities mirror those of any user in a multi-tenant setup.

Our model intentionally limits assumptions about the attacker's capabilities, diverging from conventional research that often relies on stringent conditions for executing side-channel attacks. We focus on real-world applicability and the potential for information extraction under commonly available user privileges.

\begin{figure}[t]
    \centering
    \includegraphics[width=0.8\linewidth]{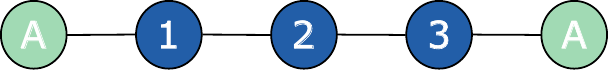}
    \caption{An example attack configuration \texttt{A123A}. `\texttt{A}' denotes malicious attacker qubits, and the numbers indicate victim qubits.}
    \label{fig:config}
\end{figure}

\begin{figure*}[htbp]
    \centering
    \begin{subfigure}[b]{0.49\textwidth} 
        \centering
        \includegraphics[width=\textwidth]{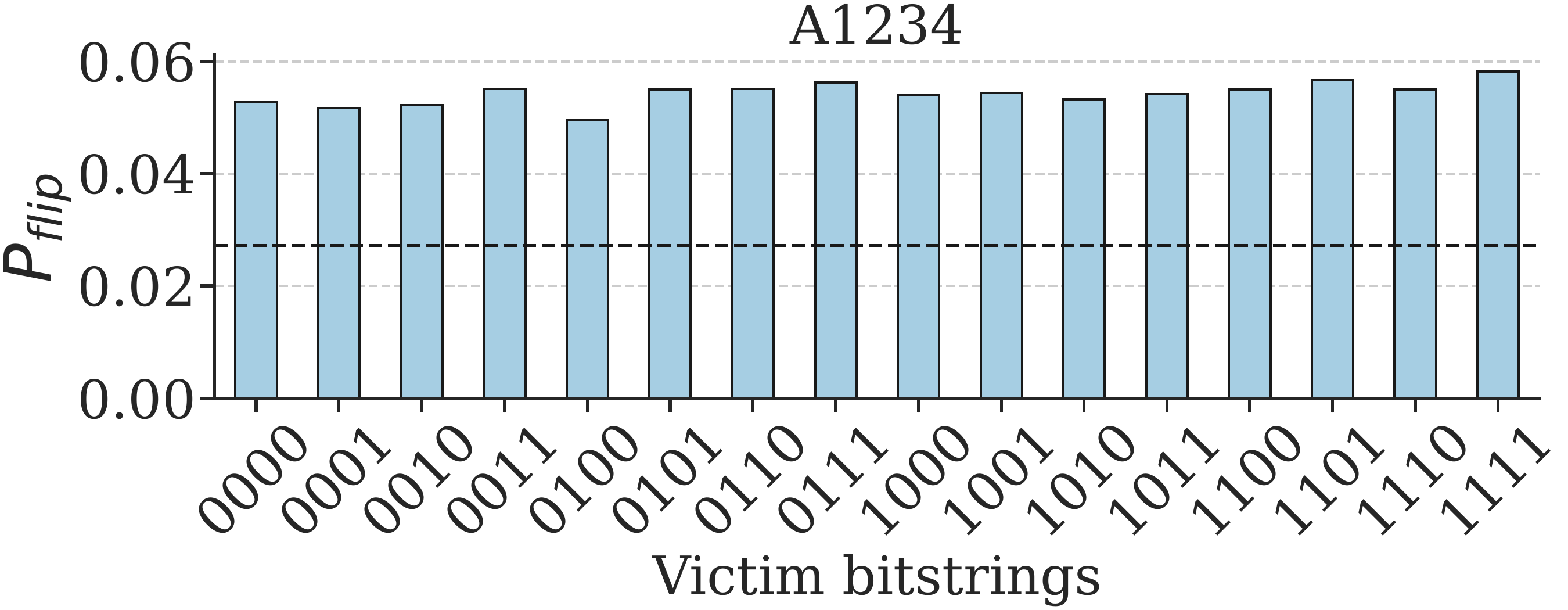}
        \caption{}
        \label{fig:sub1}
    \end{subfigure}
    \hfill
    \begin{subfigure}[b]{0.49\textwidth}
        \centering
        \includegraphics[width=\textwidth]{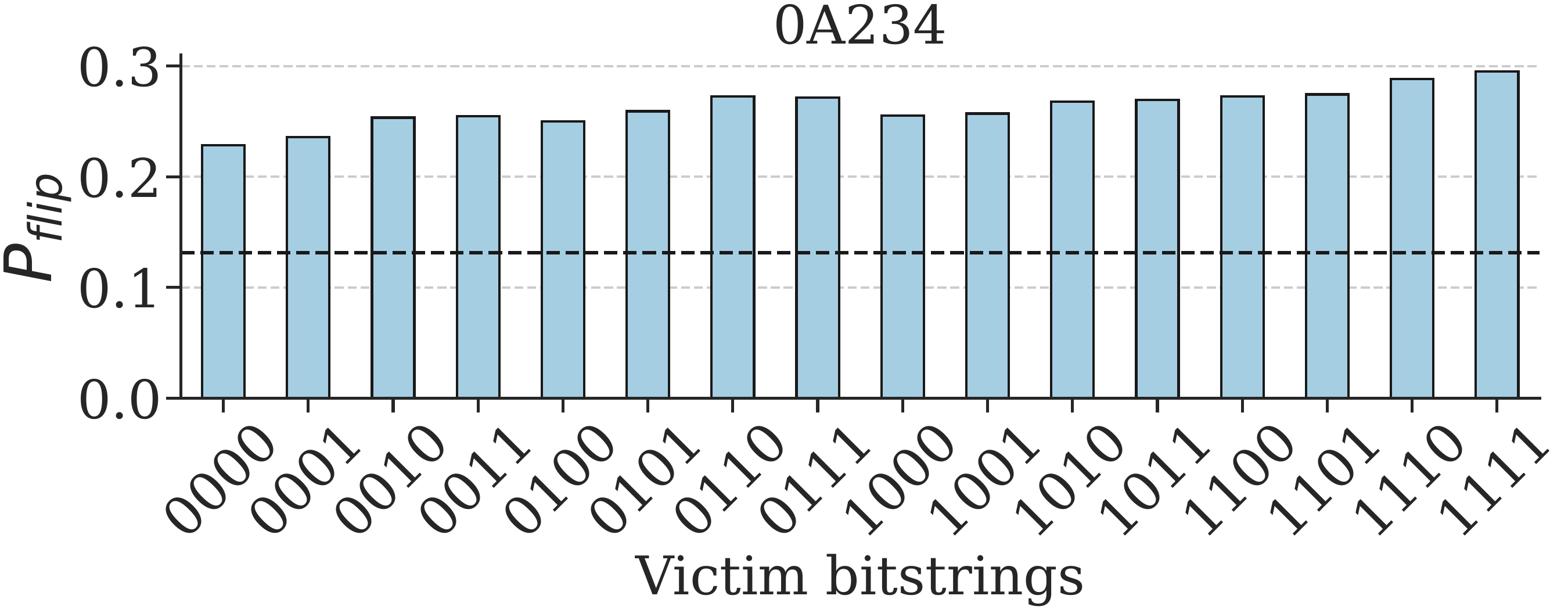}
        \caption{}
        \label{fig:sub2}
    \end{subfigure}
    \\
    \begin{subfigure}[b]{0.49\textwidth}
        \centering
        \includegraphics[width=\textwidth]{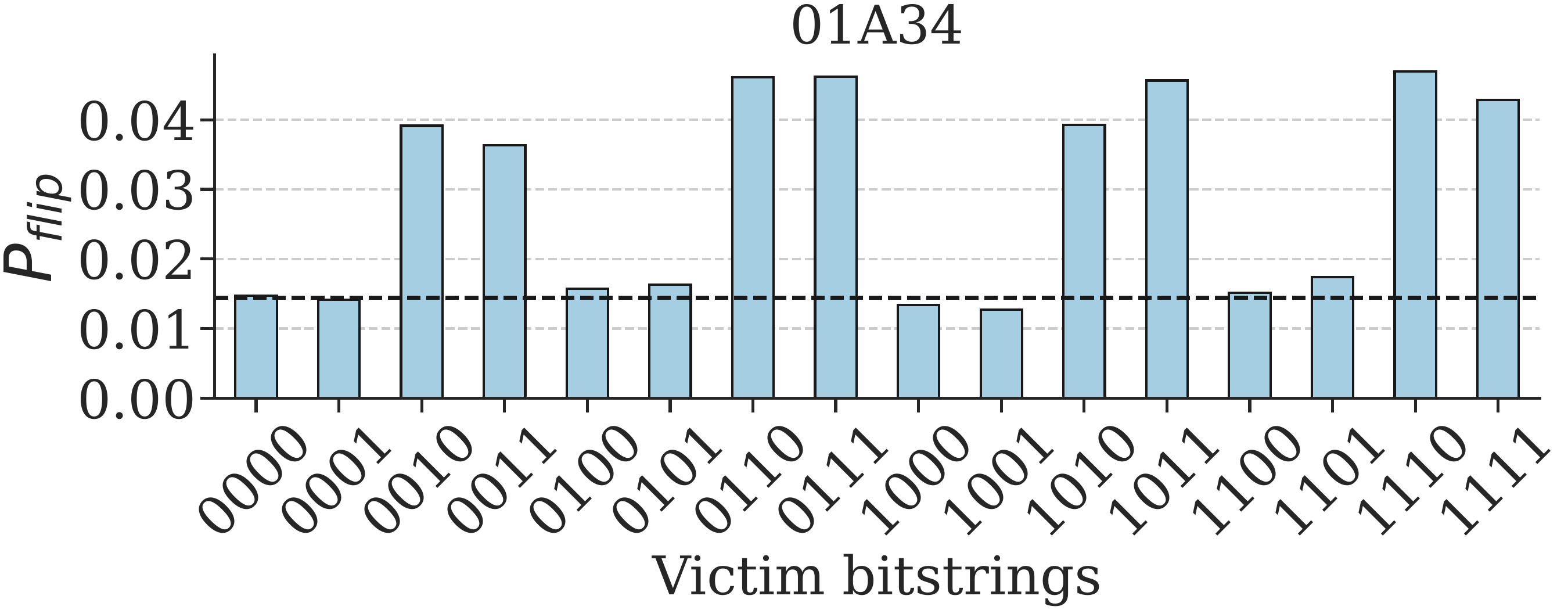}
        \caption{}
        \label{fig:sub3}
    \end{subfigure}
    \hfill
    \begin{subfigure}[b]{0.49\textwidth}
        \centering
        \includegraphics[width=\textwidth]{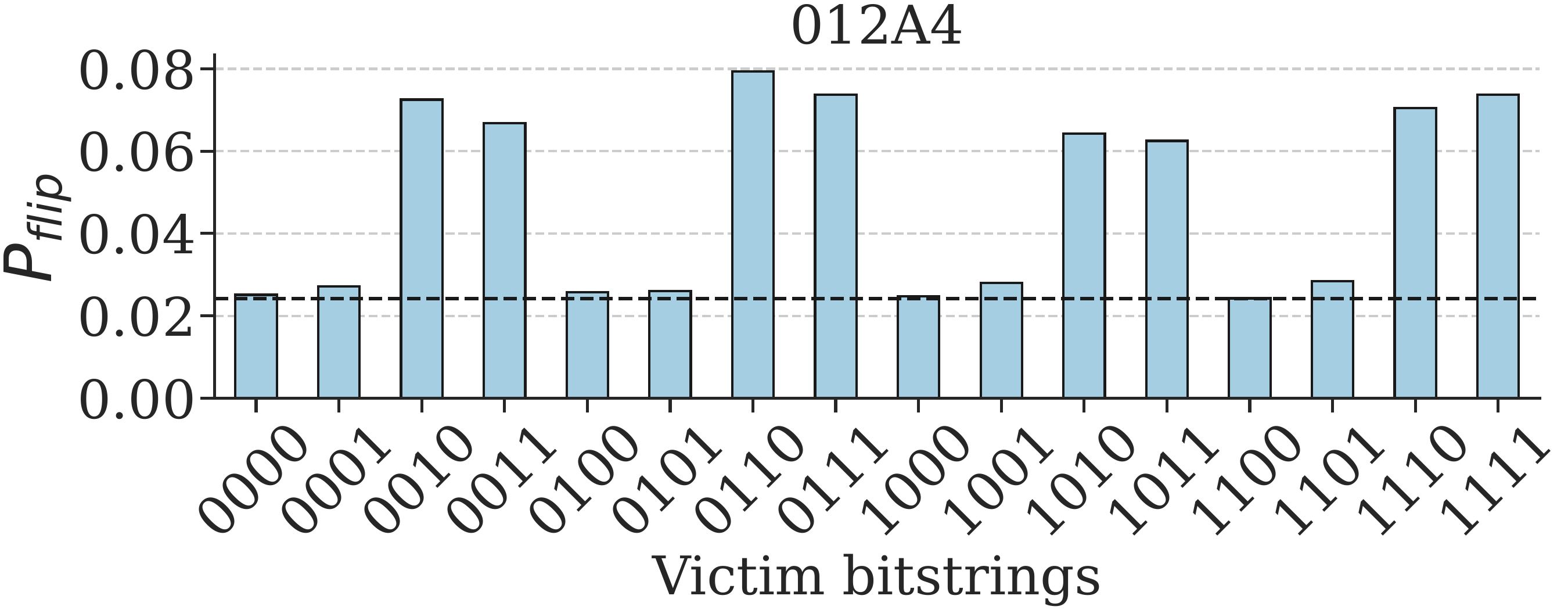}
        \caption{}
        \label{fig:sub4}
    \end{subfigure}
    \\
    \begin{subfigure}[b]{0.49\textwidth}
        \centering
        \includegraphics[width=\textwidth]{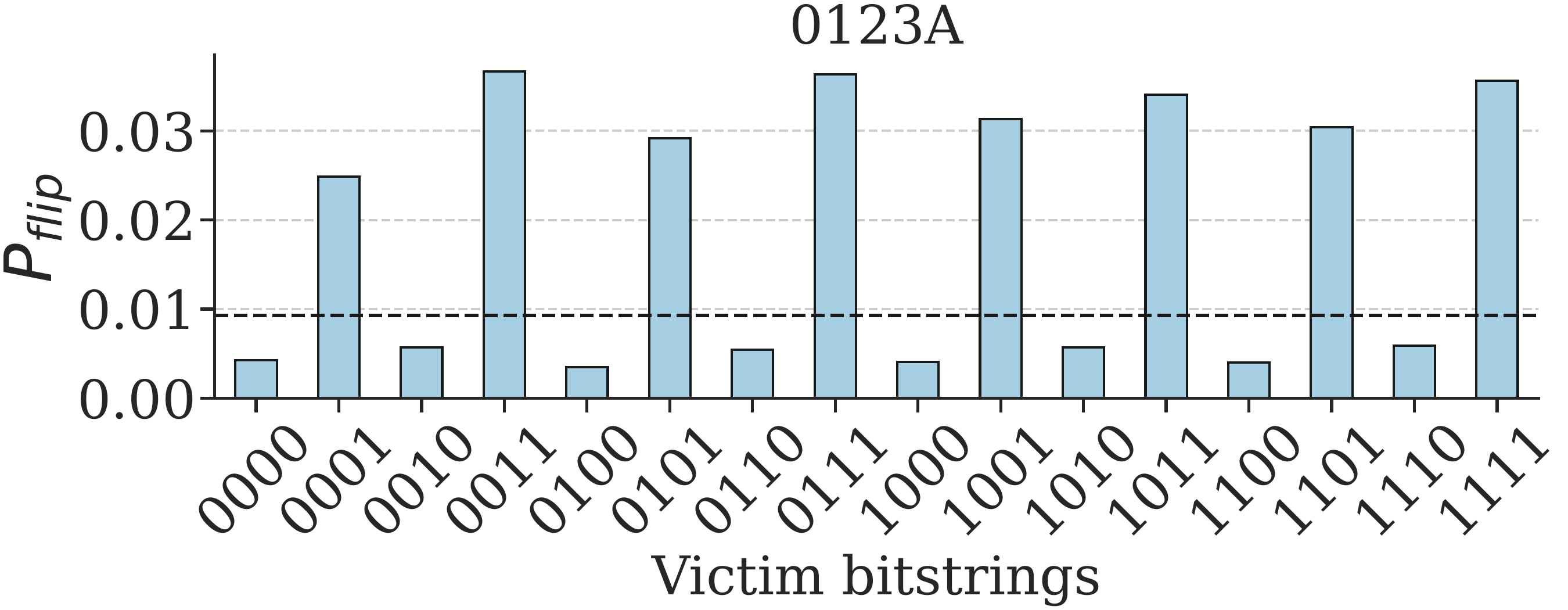}
        \caption{}
        \label{fig:sub5}
    \end{subfigure}
    \hfill
    \begin{subfigure}[b]{0.49\textwidth}
        \centering
        \includegraphics[width=\textwidth]{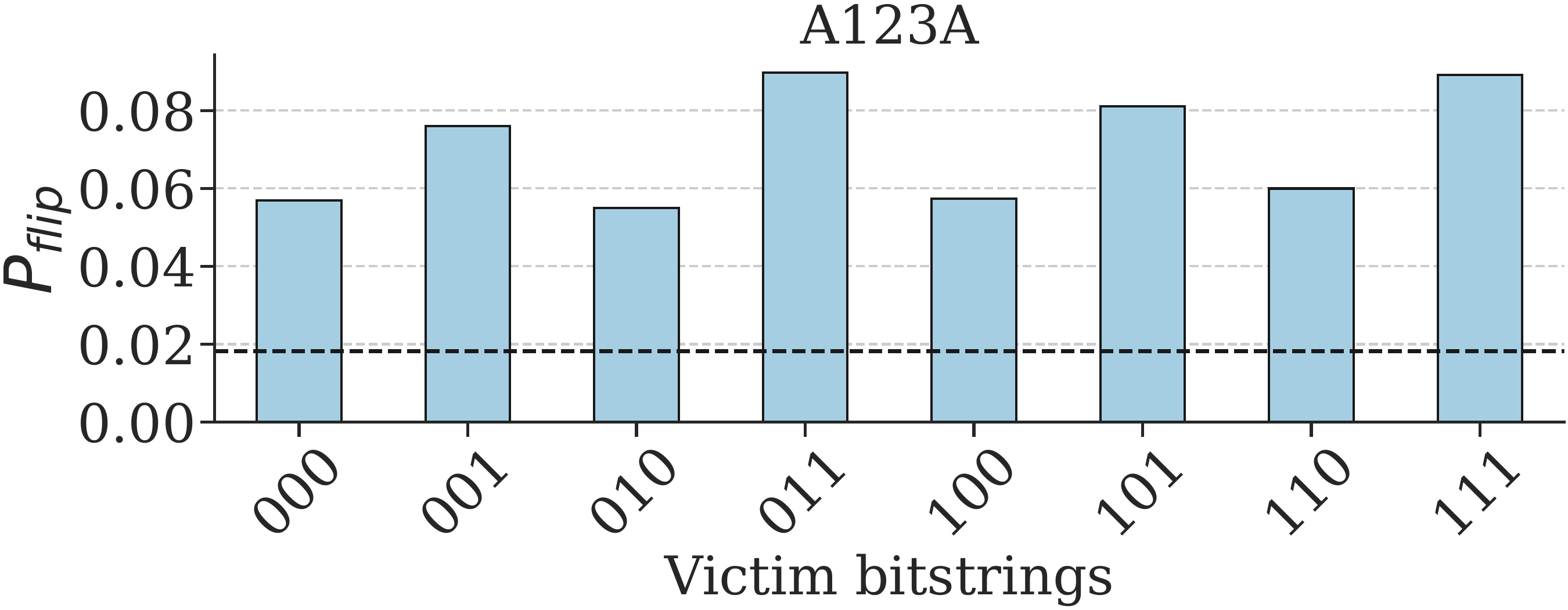}
        \caption{}
        \label{fig:sub6}
    \end{subfigure}
    \\
    \begin{subfigure}[b]{0.49\textwidth}
        \centering
        \includegraphics[width=\textwidth]{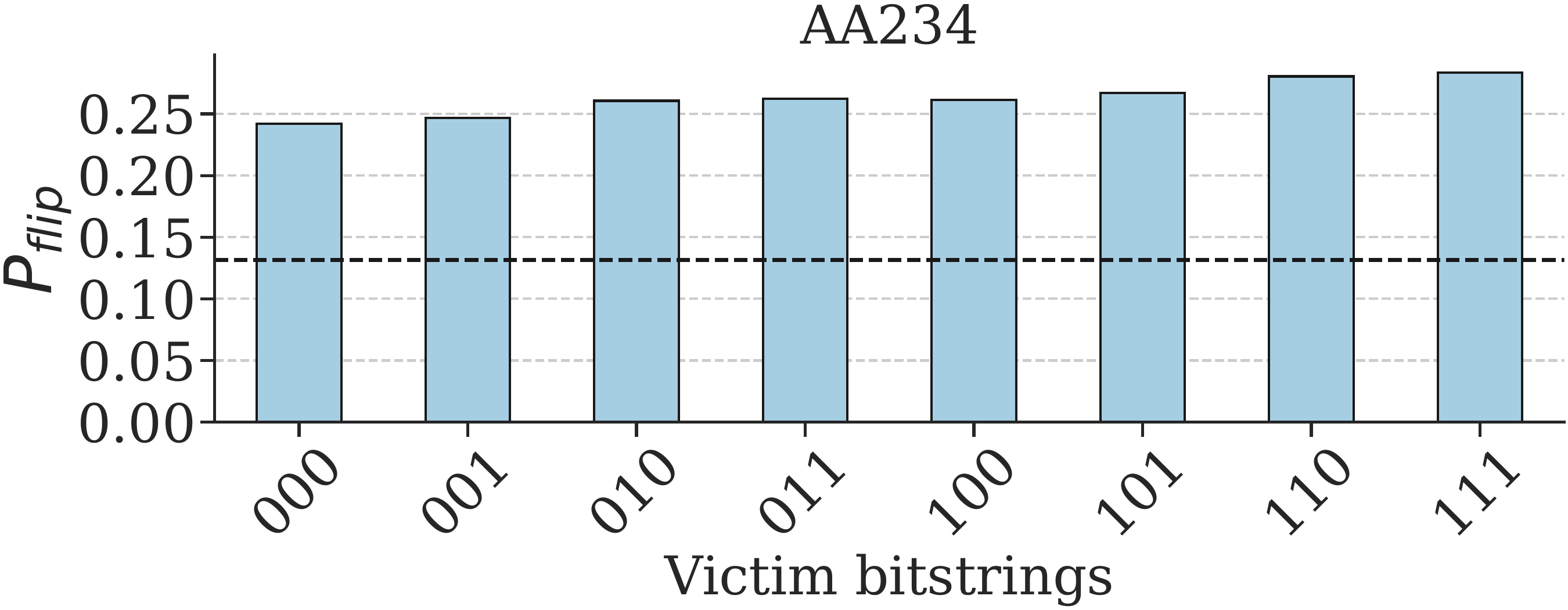}
        \caption{}
        \label{fig:sub7}
    \end{subfigure}
    \hfill
    \begin{subfigure}[b]{0.49\textwidth}
        \centering
        \includegraphics[width=\textwidth]{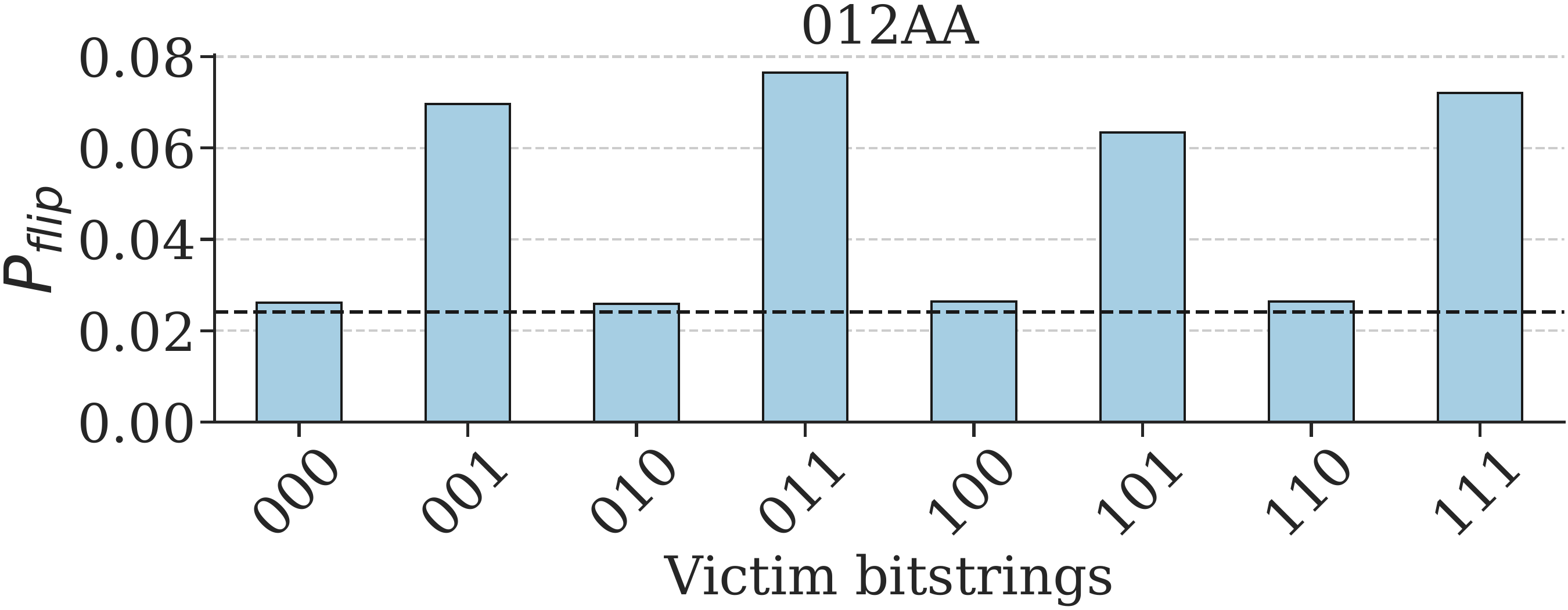}
        \caption{}
        \label{fig:sub8}
    \end{subfigure}
    \\
    \begin{subfigure}[b]{\textwidth}
        \centering
        \includegraphics[width=\textwidth]{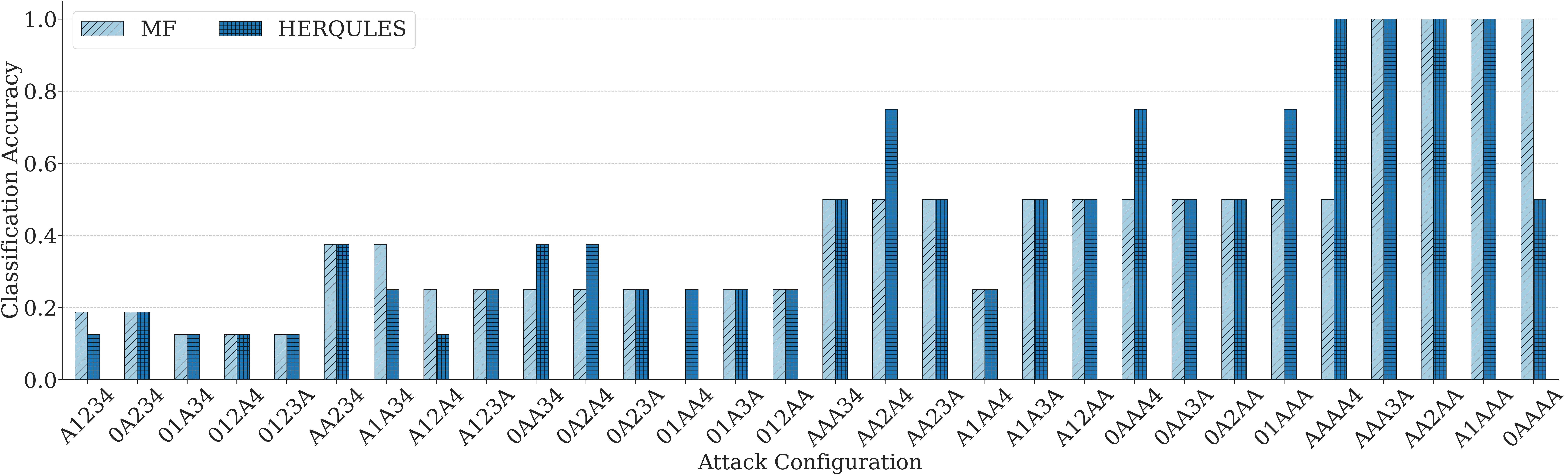}
        \caption{}
        \label{fig:acc}
    \end{subfigure}
    \caption{(a) - (h) $P_{flip}$ for different attack configurations -- the horizontal dashed line represents the average excitation error rate for the qubits being used as the attacker ($A$); (i) Classification accuracy of a simple SVM classifier predicting the victim bit-strings given the attack configuration and $P_{flip}$ for different readout discriminators (\texttt{MF}: Standard Matched Filter~\cite{PhysRevA.91.022118}; \texttt{HERQULES}:~\cite{Maurya2023}).}
    \label{fig:pflip}
\end{figure*}
\section{Evaluations} 

\subsection{Experimental Setup}

\subsubsection{Readout Dataset}

Our evaluations use the dataset generated in Reference~\cite{Lienhard2022, Maurya2023}, which consists of readout traces of five frequency-multiplexed superconducting qubits. Table~\ref{tab:resonator_freq} shows the frequency separation between the readout resonators of all five qubits. All readout traces have 500 samples each of the Inphase (I) and Quadrature (Q) components sampled at a rate of 500 MHz to correspond to a readout duration of 1$\mu$s. We use the readout discriminator results of 35,000 shots per basis state to infer the states of the five qubits. The remaining 15,000 shots in the dataset are used for training the readout discriminators.

\begin{table}[ht]
\begin{center}
\begin{small}
\caption{Resonator frequencies for 5 frequency-multiplexed qubits.}
\label{tab:resonator_freq}
\setlength{\tabcolsep}{0.15cm} 
\renewcommand{\arraystretch}{1.1}
\vspace{-0.05in}
\scalebox{0.99}{
\begin{tabular}{cccccc}
\hline \hline
Resonators            & $R_1$ & $R_2$ & $R_3$ & $R_4$ & $R_5$ \\
$\omega_R/2\pi$ (GHz) & 7.06  & 7.10  & 7.15  & 7.20  & 7.25  \\ \hline \hline
\end{tabular}
}
\end{small}
\vspace{-0.15in}
\end{center}
\end{table}

\subsubsection{Readout Discriminators}
We use the predictions made by two readout discriminators -- a standard Matched Filter (\texttt{MF}) discriminator and a neural network-based discriminator presented in~\cite{Maurya2023} (\texttt{HERQULES}). \texttt{MF} has a cumulative readout accuracy of 89.2\% and \texttt{HERQULES} has a cumulative readout accuracy of 92.7\%. Furthermore, \texttt{HERQULES} is nearly 3.5x better than \texttt{MF} in reducing correlated errors due to crosstalk~\cite{Maurya2023}. 
In this study, we do not alter the design or training of the readout discriminators -- the predictions made by the discriminators are used to study crosstalk related information leakage. 

\subsubsection{Attack Configurations}
Fig.~\ref{fig:config} shows an example attack configuration -- the qubits marked with $A$ represent the attacker qubits that are snooping on the victim qubits (annotated as numbers).

\subsubsection{Quantifying Information Leakage due to Crosstalk}
To quantify how crosstalk can leak information, we use the quantity $P_{flip}$ to determine the bit-flips from state `0' to state `1' experienced by the attacker qubits due to the state of the victim qubits.
We choose a $0\rightarrow1$ transition because unintended qubit excitations during the experiment are far less likely to occur than qubit relaxations ($1\rightarrow0$). 
Equation~\ref{eqn:pflip} shows the definition of $P_{flip}$ for an attacker qubit \texttt{A} prepared in state `0' but measured in state `1' when the victim qubits have a bit-string $V$. 

\begin{equation}
\label{eqn:pflip}
    P_{flip}(A) = Pr(A_{meas} = 1 | A_{prep} = 0, V)
\end{equation}

\subsubsection{Predicting Victim Bit-strings}
We also train a simple linear SVM to predict the state of all victim bit-strings, given the location of the attacker qubits and the value of $P_{flip}$. We train the SVM on 5,000 samples per basis state ($2^5$ total basis states for five qubits) and determine the classification accuracy on 30,000 samples per basis state.

\subsection{Research Questions}
We aim to answer the following research questions:

\begin{itemize}
    \item Can information leak due to readout crosstalk?
    \item Can the leaked information be used to predict the state of the victim qubits?
\end{itemize}

\subsection{$P_{flip}$ for Different Configurations}
$P_{flip}$ depends on the final measurement outcome after state discrimination and the kind of readout discriminator. The panels in Fig.~\ref{fig:pflip}(a)-(h) show the measured $P_{flip}$ for different attack configurations and victim bit-strings when using the \texttt{MF} readout discriminator. The horizontal dashed line represents the average excitation probability for the attacker qubits. Figs.~\ref{fig:pflip}(c), (d), (e), (f), (h) show configurations where $P_{flip}$ is higher for some bit-strings and above than the average excitation probability. In Figs.~\ref{fig:pflip}(a), (b), (g), the magnitude of $P_{flip}$ is consistently above the average excitation probability, which makes it harder to predict the states of the victim qubits. Note that $P_{flip}$ is significantly higher when qubit 2 is the attacker qubit due to its lower readout accuracy ($\sim$72\%~\cite{Lienhard2022}). 

The results presented in Fig.~\ref{fig:pflip}(a)-(h) show that for some attack configurations, some victim bit-strings can be predicted by simply measuring $P_{flip}$. With a single attacker qubit, such as shown in Fig.~\ref{fig:pflip}(e), the victim bit-strings can be bucketed depending on the magnitude of $P_{flip}$. Thus, rather than a probability of $1/16$ for predicting the victim bit-string correctly without any crosstalk information, the probability can be 2-4$\times$ higher if $P_{flip}$ is determined.

\subsection{Prediction Accuracy}
Having shown that $P_{flip}$ is higher for some victim bit-strings in some attack configurations, we now evaluate whether this information can predict the victim bit-strings. Fig.~\ref{fig:acc} shows the classifier accuracy of a simple SVM for different attack configurations. As expected, the accuracy increases as the number of attacker qubits increases. Furthermore, the discrimination results of \texttt{HERQULES} enable better classification accuracies than \texttt{MF} for some attack configurations -- this is an important observation because \texttt{HERQULES} is a significantly better discriminator in terms of overall accuracy and crosstalk mitigation~\cite{Maurya2023}. Therefore, accessing side-channel information can become more accessible as we improve the state discrimination accuracy by explicit crosstalk detection and suppression. Note that when the first qubit is the attacker qubit (cases $A1234$, $A1A34$, $A12A4$), the classification accuracy can be higher for the \texttt{MF} discriminator because that qubit incurs very few correlated errors due to crosstalk, which are effectively eliminated by \texttt{HERQULES}. For the same reason, the classification accuracy with \texttt{HERQULES} is lower for $0AAAA$.






\section{Discussion} 

\subsection{Defense using Circuit Sandboxing}

\begin{figure}[t]
    \centering
\includegraphics[width=0.75\linewidth]{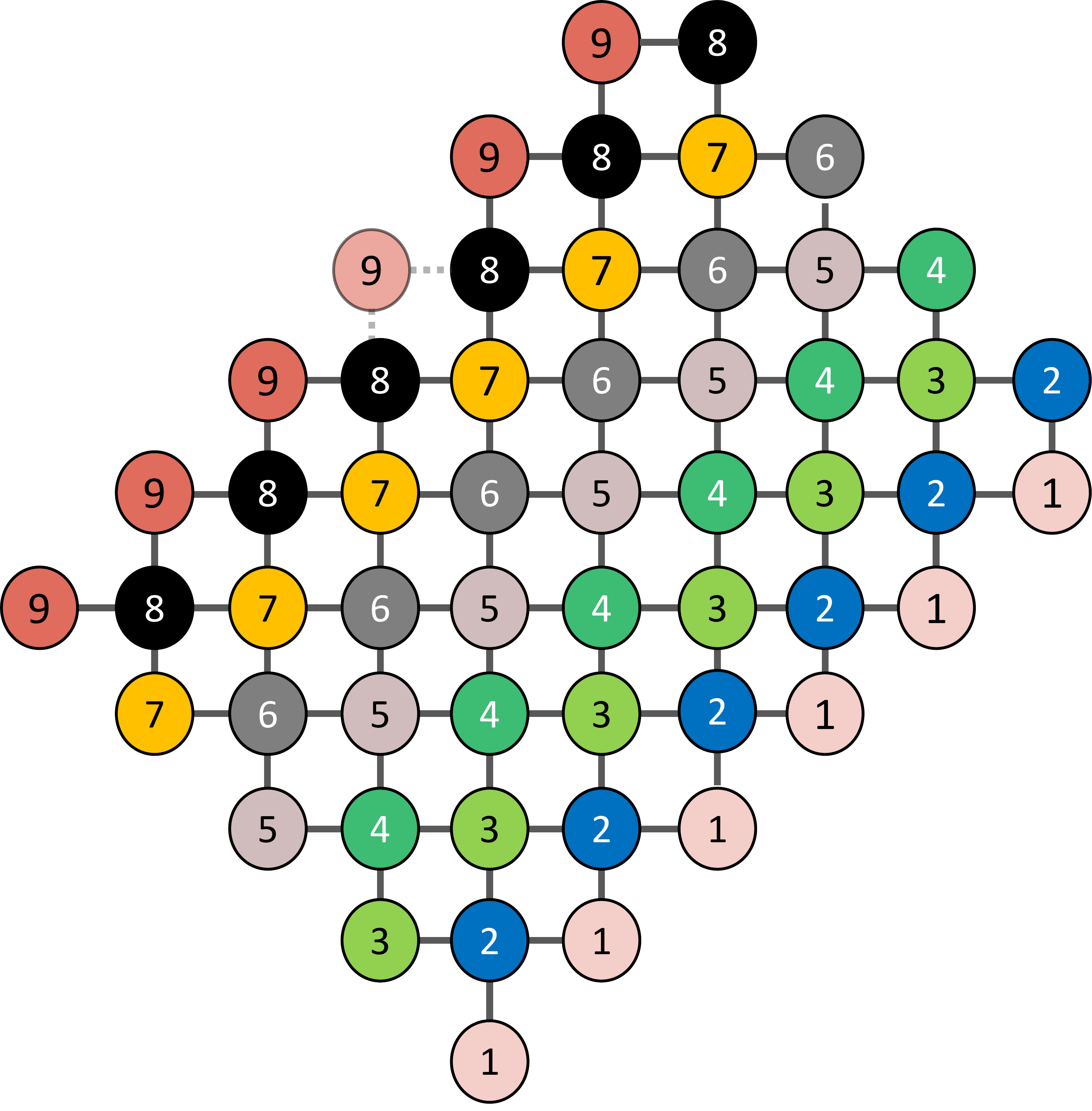}
    \caption{Presumed map of qubit groups for frequency-multiplexed readout in the Sycamore quantum processor~\cite{quantumsupremacy_nature_2019}.}
    \label{fig:map}
    \vspace{-0.1in}
\end{figure}

To mitigate side channel leakage, it is advisable not to distribute qubits with shared readout feedlines among multiple users. However, in an architecture with frequency-multiplexed readout, where five to six qubits are linked to one feedline, partitioning at the feedline boundary can lead to fixed-granularity assignment of qubits, resulting in underutilization. For instance, Fig.~\ref{fig:map} shows the presumed qubit readout group for Google's Sycamore processor, where qubits within a group share a feedline. The 54 qubits in the Sycamore system are organized into nine distinct groups for frequency-multiplexed readout, each containing six qubits. Within each group, individual qubits are connected to unique readout resonators, all coupled to a common bandpass Purcell filter. This arrangement facilitates simultaneous readout of all qubits within a group. The readout process utilizes frequency-domain multiplexing, where the collective probe signal comprises overlapping probe signals at the resonant frequencies of each readout resonator. The resonators can be probed in parallel through this method, allowing for efficient and concurrent data acquisition from the six qubits.

If five qubits are connected to a single feedline, users would be assigned an integer multiple of five to six qubits. Moreover, suppose the qubits read out through a shared feedline are not connected via qubit couplers, essential for performing two-qubit gates~\cite{Fei2018_Coupler,quantumsupremacy_nature_2019}. In that case, the underutilization can be even more significant. When allocating qubits to users, each user needs to be allocated a connected graph of qubits so that any two qubits can entangle while running a quantum circuit. 



\subsection{Defense using Circuit Mapping Randomization}

To enhance security and prevent malicious use of side channels, we can randomize the location mapping of users on quantum computers. This approach, commonly used in classical cloud services, prevents the same two users from being frequently positioned close together. Specifically, in the quantum context, cloud providers can execute `$N$' circuit shots that are not on the same set of qubits every time. Instead, they select different qubits for each execution or group of executions, thereby diversifying the physical qubits used and minimizing predictable patterns that could be exploited. Prior work has used an ensemble of qubit mapping to reduce correlated errors and enhance the fidelity of NISQ circuits~\cite{tannu2019ensemble}. 

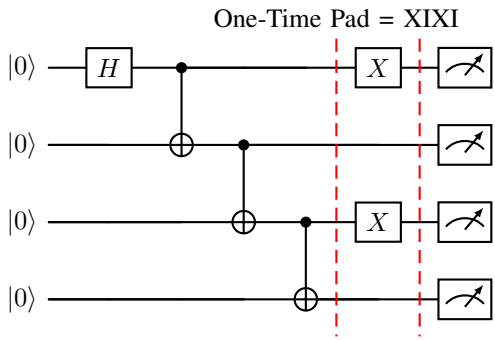
\begin{figure}[t]
    \centering
\begin{quantikz}
   \lstick{\ket{0}} & \gate{H} & \ctrl{1} & \qw      & \qw     \slice{One-Time Pad = XIXI} & \gate{X} \slice{} & \meter{} \\
   \lstick{\ket{0}} & \qw      & \targ{}  & \ctrl{1} & \qw      & \qw & \meter{} \\
   \lstick{\ket{0}} & \qw      & \qw      & \targ{}  & \ctrl{1} & \gate{X} & \meter{} \\
    \lstick{\ket{0}} & \qw      & \qw      & \qw      & \targ{}  & \qw & \meter{}
\end{quantikz}

    \caption{GHZ circuit with output scrambling using a one-time pad $X \otimes I\otimes X \otimes I$.}
    \label{fig:GHZexample}
\end{figure}

Implementing randomized mapping to enhance security is practical but introduces complexities in resource management. Firstly, it requires user circuits to be compiled for various subsets of physical qubit mappings without significantly increasing the number of gates or the overall circuit complexity. Additionally, if multiple users with malicious intent collude, they could manipulate job allocation to achieve co-location, bypassing the security measure. Another challenge is the impact on error mitigation strategies. These techniques often depend on understanding the specific noise characteristics of quantum devices to correct errors effectively. Constantly changing qubit mappings can disrupt this understanding, potentially reducing the effectiveness of error suppression and correction methods.


\subsection{Defense using Scrambling}
To protect user data against readout crosstalk, we can scramble output data by inserting X-gates on randomly selected qubits by using these strings of X-gates as a one-time pad. This one-time pad is only known to the user, so the data is hidden from the attacker. While the output state is scrambled and hidden from the attacker, it can be easily unscrambled by XORing the output bit strings with the one-time pad running a classical post-processing step. To illustrate the scrambling and unscrambling, consider Fig.~\ref{fig:GHZexample}, where a four qubit circuit that produces a GHZ state is protected by adding a randomly selected one-time pad of $X \otimes I\otimes X \otimes I$ to scramble the output state. The measured output can be XORed with 1010 during classical post-processing to unscramble. Although effective in hiding the exact output state, simple scrambling patterns outlined in this section can still leak the degree of entanglement. Therefore, it is essential to understand the security risks of multi-tenancy and adopt an execution model tailored to security needs.



\section{Conclusion}

As quantum processors based on superconducting qubits scale up, frequency-multiplexed readout will be increasingly important in reducing the resource requirements for qubit readout. However, this frequency-multiplexing comes at the cost of correlated errors due to increased readout crosstalk. While prior works have focused on reducing the errors caused by readout crosstalk, we show how correlated errors due to readout crosstalk can be exploited to predict the state of victim qubits in a multi-programmed quantum computer. Our evaluations show that readout crosstalk causes attacker qubits to flip from $0\rightarrow1$ depending on the state of the victim qubits. These state-dependent bit-flips in the attacker qubits can be used to train classifiers that can predict the entire victim bit-string. Scrambling the to-be-measured quantum information and enabling the target user to unscramble the information classically may be a potential path to protect users from attacks.

\bibliographystyle{ieeetr}
\bibliography{refs}

\end{document}